\documentstyle[12pt]{article}
\newcommand{\CMP}[1]{{\em Commun. Math. Phys.} {\bf {#1}}}
\newcommand{\JMP}[1]{{\em J.~Math. Phys.} {\bf {#1}}}

\newcommand{\Ann}[1]{{\em Ann. Phys.} {\bf {#1}}}
\newcommand{\sss}[1]{{\bf {#1}:}}

\newcommand{\restr}[1]{\left.\right|_{#1}}

\newcommand{\eq}{\begin{equation}}
\newcommand{\eqend}{\end{equation}}
\newcommand{\eqa}{\begin{eqnarray}}
\newcommand{\neqa}{\begin{eqnarray*}}
\newcommand{\eqaend}{\end{eqnarray}}
\newcommand{\neqaend}{\end{eqnarray*}}
\newcommand{\nonu}{\nonumber \\ \nopagebreak}
\newcommand{\bma}[1]{\begin{array}{#1}}
\newcommand{\ema}{\end{array}}
\newcommand{\bc}{\begin{center}}
\newcommand{\ec}{\end{center}}

\newcommand{\Ref}[1]{(\ref{#1})}

\newcommand{\lrar}{\longrightarrow}

\newcommand{\Biar}{\Longleftrightarrow}

\newcommand{\dd}{\mbox{${\rm d}$}}

\newcommand{\ee}[1]{\mbox{{\rm e}}^{#1}}
\newcommand{\ii}{{\rm i}}
\newcommand{\OO}{{\rm O}}

\renewcommand{\det}[1]{\mbox{{\rm det}} ({#1})}
\newcommand{\detp}[1]{\mbox{{\rm det}'} ({#1})}

\newcommand{\dif}[1]{{\rm deg}({#1})}

\renewcommand{\vec}[1]{\mbox{\boldmath ${#1}$}}

\newcommand{\lam}{\lambda}
\renewcommand{\phi}{\varphi}
\newcommand{\sig}{\sigma}
\newcommand{\del}{\delta}

\newcommand{\Om}{\Omega}

\newcommand{\eps}{\varepsilon}

\newcommand{\sign}{{\rm sign}}

\newcommand{\R}{{\sf I} \! {\sf R}}
\newcommand{\C}{{\sf C} \! \! \! {\sf I}\:}

\newcommand{\N}{{\sf I} \! {\sf N}}
\newcommand{\Z}{{\sf Z} \! \! {\sf Z}}

\newcommand{\f}{\frac}

\newcommand{\cD}{{\cal D}}

\newcommand{\cP}{{\cal P}}

\newcommand{\cF}{{\cal F}}
\newcommand{\cE}{{\cal E}}

\newcommand{\cU}{{\cal U}}

\newcommand{\hGam}[1]{\hat{\Gamma}  ({#1})}          
\newcommand{\dhgam}[1]{{\rm d}\hat{\Gamma} ({#1})}  

\newcommand{\ccr}[2]{{[} {#1},{#2} {]} }        
\newcommand{\scr}[2]{{[} {#1},{#2} {]}_X }        

\newcommand{\produ}[2]{( {#1},{#2} ) }          
\newcommand{\Produ}[2]{< {#1},{#2} > }          

\newcommand{\tra}[1]{{\rm tr} ({#1})}          
\newcommand{\norm}[1]{|{#1}| }                 
\newcommand{\Norm}[1]{|\! | {#1} |\! |}        
\newcommand{\NORM}[1]{|\! |\! |{#1} |\! |\! |} 

\newcommand{\Df}{\cD_X^f(h)}           
\newcommand{\BO}{B(h)}            
\newcommand{\TR}{B_{1} (h)}       
\newcommand{\HS}{B_{2} (h)}       

\newcommand{\gX}{\vec{g}_{X}(h)}
\newcommand{\gXu}{\vec{g}^{T_-}_{X,2}(h;H)}
\newcommand{\GB}{\vec{G}_{B}(h)}
\newcommand{\GF}{\vec{G}_{F}(h)}
\newcommand{\GX}{\vec{G}_{X}(h)}
\newcommand{\GBn}{\vec{G}_{B}^{(0)}(h)}
\newcommand{\GFn}{\vec{G}_{F}^{(0)}(h)}
\newcommand{\GXn}{\vec{G}_{X}^{(0)}(h)}
\newcommand{\GXnn}{\vec{G}_{X,0}(h)}

\newcounter{saveeqn}
\newcounter{App} 

\renewcommand{\include}{\input}
\author{Edwin Langmann
\thanks{Erwin Schr\"odinger-fellow, supported by the ``Fonds zur
F\"orderung der Wissenschaftlichen Forschung in \"Osterreich'' under
the contract Nr.~J0789-PHY}
\\ Department of Physics\\ The University of British Columbia\\
V6T 1Z1 Vancouver, B.C., Canada}
\title{Cocycles for Boson and Fermion Bogoliubov Transformations}
\date{August 7, 1992}

\begin{document}
\thispagestyle{empty}
\maketitle
\vspace*{30.0mm}
\thispagestyle{empty}
\begin{abstract}
We discuss unitarily implementable Bogoliubov transformations for
charged, relativistic bos\-ons and fermions, and we derive explicit
formulas for the 2-cocycles appearing in the group product of their
implementers. In the fermion case this provides a simple field
theoretic derivation of the well-known cocycle of the group of unitary
operators on a Hilbert space modeled on the Hilbert Schmidt class and
closely related to the loop groups. In the boson case the cocycle is
obtained for a similar group of pseudo-unitary (symplectic) operators.
We also give formulas for the phases of one-parameter groups of
implementers and, more generally, families of implementers which are
unitary propagators with parameter dependent generators.
\end{abstract}

\section{Introduction}
\label{sec0}
The formalism for quantizing relativistic fermions in external fields
is not only essential for quantum field theory, but it plays also a
crucial role in the representation theory of the affine Kac-Moody
algebras and the Virasoro algebra \cite{GO}. Indeed, this connection
has led to a most fruitful interplay between physics and mathematics
(see e.g.\ \cite{VMP}).

The geometric approach to this subject by means of determinant bundles
over infinite dimensional Grassmannians \cite{PS,M} seems to be
preferred by mathematicians but is quite abstract and different from
the physicists' tradition.
There is, however, another rigorous approach in the spirit and close
to quantum field theory, namely the theory of {\em quasi-free second
quantization} (QFSQ) of fermions. Lundberg \cite{L} was probably the
first who formulated its abstract framework in an elegant and concise
way, and he used it to construct in general abstract current algebras
providing (by restriction) representations of the affine Kac-Moody
algebras and the Virasoro algebra \cite{CR}. Later on this formalism
was worked out in all mathematical detail by Carey and Ruijsenaars
\cite{CR}.

Besides its conceptual simplicity, QFSQ of fermions has another
advantage, namely it has a natural boson counterpart (which to our
knowledge is not the case for the Grassmannian approach \cite{PS,M}).
Ruijsenaars \cite{R1} in his comprehensive work on Bogoliubov
transformations of charged, relativistic particles made clear and
exploited the formal analogy of bosons and fermions, and he was able
to derive most of the corresponding formulas for the two cases in a
parallel way. Though very transparent and simple, Ruijsenaar's {\em
QFSQ of bosons} did not become very popular probably due to the fact
that it deviates substantially from the traditional approach
to boson quantum
field theory (which is usually formulated in terms of Weyl operators
(= exponentiated fields) \cite{BR2} rather than the fields itself).

Though in their extensiv work on QFSQ Carey and Ruijsenaars \cite{CR}
restricted themselves to the fermion case, it is rather
straightforward to derive most of their results for the boson case as
well \cite{Runpub}, and, moreover, to develop a $\Z_2$-graded
formalism --- a super-version of QFSQ --- \cite{Runpub,GL1,GL2}
comprising the boson and the fermion case and extending these in a
non-trivial way. Especially, the resulting current super algebras
naturally provide representations of the $\Z_2$-graded extensions of
the affine Kac-Moody algebras and the Virasoro algebra \cite{GL1,GL2}.

The implementers of Bogoliubov transformations \cite{R1} are an
`integrated version' of the currents referred to above, i.e.\,
the abstract current algebra provides the Lie algebra of the Lie group
generated by these implementers. It is
well-known that the essential, non-trivial aspect of the current
algebras in the occurance of a Schwinger term \cite{CR,GL1} which ---
from a mathematical point of view --- is a non-trivial Lie algebra
2-cocycle. On the group level, this corresponds to a non-trivial Lie
group 2-cocycle arising in the product relations of the implementers
\cite{PS,M}.

In this paper we prove the explicit formulas for these group
2-cocycles in the standard phase convention \cite{R1} by a simple,
direct calculation. Moreover, we derive a general formula for the
phase relating the one-parameter group of implementers obtained by
exponentiating a current (via Stone's theorem
\cite{RS1}) to the implementer given by the general formula in
\cite{R1}, and we generalize  this to unitary propagators \cite{RS2}
generated by `time' dependent currents.

Our results can be regarded as a supplement to \cite{R1}. However (in
contrast to \cite{R1}), our proofs are completely parallel in the
boson and the fermion case: we introduce a symbol $X\in\{B,F\}$ which
is equal to $B$ in the boson and to $F$ in the fermion case, and all our
formulas and arguments are given in terms of this variable $X$. To
this aim we introduce the symbols
\eqa
\dif B \equiv 1, \quad \dif F \equiv 0\nonu
\eps_B\equiv +1, \quad \eps_F\equiv -1,
\eqaend
i.e.\ $\eps_X=(-)^{\dif X}$, and
\eq
\scr{a}{b}\equiv ab-\eps_X ba,
\eqend
i.e.\ $\ccr{\cdot}{\cdot}_B$ is the commutator and
$\ccr{\cdot}{\cdot}_F$ is the anticommutator.

In the fermion case we were not able to obtain the result in the same
generality as Ruijsenaars \cite{R1}, but our formulas are restricted
to some neighborhood of the identity containing the topologically
trivial Bogoliubov transformations leaving the particle number and the
charge unchanged \cite{R1,CR}. Moreover, our formula for the group
2-cocycle is well-known in the fermion case \cite{PS}. As it plays an
essential role in the theory of the loop groups \cite{PS} with all its
applications \cite{VMP}, we hope that our alternative proof is
nevertheless useful.  To our knowledge, the formula in the boson case
is new. Moreover, these 2-cocycles play a crucial role for the
construction of current algebras for bosons and fermions in
$(3+1)$-dimensions (arising from Bogoliubov transformations which are
not unitarily implementable but require some additional ``wave
function renormalization'') given recently by the author \cite{EL2}
(the formulas were used in \cite{EL2} without proof).

The plan of this paper is as follows.
In the next Sect.\ we introduce our notation and summarize the facts
we need about Bogoliubov transformations and currents algebras within the
framework of QFSQ of bosons and fermions.
Our results are presented in Sect.\ \ref{sec2} and their proofs are
given in Sect. \ref{sec3}. We conclude with a few remarks in Sect.\
\ref{sec4}.

\section{Preliminaries}
\label{sec1}
\sss{(a) Second Quantization} Let $h$ be a separabel Hilbert space
 and the direct sum of
two subspaces $h_+$ and $h_-$: $h=h_+\oplus h_-$, and $T_\pm$ the
orthogonal projections in $h$ onto $h_\pm$: $h_\pm = T_\pm h$.
$h_+$ can be thought of as the one-{\em particle} space and $h_-$ as
the one-{\em antiparticle} space. We write $\BO$, $\HS$, and $\TR$ for
the bounded, the Hilbert-Schmidt, and the trace-class operators on $h$
\cite{RS1}, respectively, and for any linear operator $A$ on $h$,
\eq
\label{a1}
A_{\eps\eps'}\equiv P_\eps A P_{\eps'} \quad \forall
\eps,\eps'\in\{+,-\}.
\eqend
We denote as $\cF_B(h)$ ($\cF_F(h)$) the boson (fermion) Fock space
over $h$ with the vacuum $\Om$ and creation and annihilation operators
$a^*(f)$ and $a^{}(f)$, $f\in h$ obeying canonical commutator
(anticommutator) relations
\eqa
\label{a1a}
\scr{a^{}(f)}{a^*(g)} &=& (f,g) \nonu
\scr{a^{}(f)}{a^{}(g)} &=& 0 \quad \forall f,g\in h
\eqaend
($(\cdot,\cdot)$ is the inner product in $h$), and
\eq
\label{a2}
a^{}(f)\Om = 0, \quad a^*(f)= a^{}(f)^* \quad \forall f\in h
\eqend
as usual \cite{BR2} ($*$ denotes the Hilbert space adjoint).
Moreover, we introduce
the particle number operator $N$
inducing a natural $\N_0$-gradation in $\cF_X(h)$
\[
\cF_X(h)=\bigoplus_{\ell = 0}^{\infty} h_X^{(\ell)}, \quad
h_X^{(\ell)} = (P_\ell - P_{\ell -1})\cF_X(h)
\]
with $h_X^{(\ell)}$ the $\ell$-particle subspace and $P_\ell$ the
orthogonal projection onto the vectors with particle number less or
equal to $\ell$ \cite{BR2}, and
\eq
\label{a3}
\Df \equiv
\left\{\eta\in\cF_X(h)| \exists \ell <\infty: P_\ell \eta = \eta \right\}
\eqend
is the set of finite particle vectors; note that $\Df$ is dense in
$\cF_X(h)$ \cite{R2}. Similar as Ruijsenaars \cite{R1}, we introduce
the field operators
\eqa
\label{a4}
\Phi^+(f) &\equiv& a^*(T_+ f) + a^{}(JT_- f) \nonu
\Phi(f) &\equiv& a^{}(T_+ f) - \eps_X a^*(JT_- f) \quad \forall f\in h
\eqaend
with $J$ a conjugation in $h$ commuting with $T_\pm$.
Then the (anti-) commutator relations \Ref{a1} result in
\eqa
\label{a7a}
\scr{\Phi^{}(f)}{\Phi^+(g)} &=& (f,g) \nonu
\scr{\Phi^{}(f)}{\Phi^{}(g)} &=& 0 \quad \forall f,g\in h,
\eqaend
and
\eq
\label{a7b}
\Phi^{}(f) \equiv \Phi^+(q_X f)^* \quad \forall f\in h
\eqend
with $q_X=T_+ -\eps_X T_-$, i.e.\
\eq
\label{a8}
q_B\equiv P_+-P_-, \quad q_F\equiv 1.
\eqend

{\bf Remark 2.1:} Note that the operators $a^{(*)}(f)$ and $\Phi^{(+)}(f)$
($f\in h$) are bounded in the fermion- but unbounded in the boson case
\cite{BR2}. However, in the later case, $\Df$ \Ref{a3} provides a
common, dense, invariant domain for all these operators due to the
estimate \cite{R1,GL2} (which trivially hold for $X=F$ as well)
\eq
\label{est}
\Norm{a^{(*)}(f)P_\ell} \leq (\ell +1)\Norm{f} \quad \forall f\in h,
\ell \in\N
\eqend
(with $\Norm{\cdot}$ we denote the operator norm and the Hilbert space
norm), and they are closed operators on $\cF_B(h)$. Hence \Ref{a7a} and
similar eqs.\ below have to be understood as relations on $\Df$.

{\bf Remark 2.2:} Note that Ruijsenaars \cite{R1} uses the field
operators $\tilde{\Phi}^*(f)=\Phi^+(f)$ and
$\tilde{\Phi}(f)=\tilde{\Phi}^*(f)^*=\Phi^{}(q_X f)$ ($f\in h$)
deviating from ours in the boson case $X=B$. Our definition is more
convenient for discussing the current algebras (see below).

\sss{(b) Bogoliubov Transformations}
Let $U$ be a closed, invertible operator on $h$. Then the
transformation
\eq
\label{a9}
\alpha_U :
\Phi^+(f)\mapsto \alpha_U(\Phi^+(f)) \equiv \Phi^+(Uf) \quad \forall
f\in h
\eqend
leaves the relations \Ref{a7a}, \Ref{a7b} invariant if and only if
\eq
\label{a10}
Uq_XU^* = U^*q_X U = q_X .
\eqend
We denote such an $U$ as {\em $X$-unitary} \footnote{note that
$F$-unitary=unitary, and what we call $B$-unitary was denoted as
{\em pseudo-unitary} in \cite{R1}
}, and $\alpha_U$ \Ref{a9} is a {\em Bogoliubov transformation} (BT).
It is called {\em unitarily implementable} if there is an unitary
operator $\hGam{U}$ on $\cF_X(h)$ such that
\eq
\label{a11}
\hGam{U}\Phi^+(f) = \Phi^+(Uf)\hGam{U} \quad \forall f\in h,
\eqend
and the well-known necessary and sufficient condition for this to be
the case is the Hilbert-Schmidt criterium \cite{R1}
\eq
\label{a12}
U_{+-}, U_{-+} \in\HS .
\eqend
(Though not completely obvious, by a little thought one can convince
oneselves that our $\hGam{U}$ is
identical with the implementer $\cU$ defined in \cite{R1}, eqs.\ (2.10)
and (2.18)).
We denote the group of all $X$-unitary operators on $h$ obeying this
condition as $\GX$. Furthermore, we introduce the set $\cU^{(0)}(h)$
of all closed, invertible operators $U$ on $h$ such that $U_{--}$ has
a bounded inverse $(U_{--})^{-1}$ on $h_-$, and
\eq
\label{a13}
\GXn\equiv \GX\cap \cU^{(0)}(h) .
\eqend

A crucial difference between the boson and the fermion
case is that \Ref{a10} and \Ref{a12} for $X=B$ (but {\em not} for
$X=F$) imply that $U\in\cU^{(0)}(h)$, hence
\eq
\label{a14}
\GBn=\GB,
\eqend
whereas there are plenty of $U\in\GF$ not contained in $\GFn$
\cite{R1}, and there is a one-to-one correspondence between $\GBn$ and
$\GFn$ showing that there are many more fermion than boson BTs: Indeed,
for $U\in\cU^{(0)}(h)$ the eq.\
\eq
\label{a15}
T_- = UT_+ - T_+Z + UT_-Z = ZT_+ - T_+ U + ZT_- U
\eqend
has a unique solution $Z\in\cU^{(0)}(h)$,
\eqa
\label{a16}
Z_{++} &=& U_{++} - U_{+-}(U_{--})^{-1}U_{-+} \nonu
Z_{+-} &=& U_{+-}(U_{--})^{-1} \nonu
Z_{-+} &=& -(U_{--})^{-1}U_{-+} \nonu
Z_{--} &=& (U_{--})^{-1},
\eqaend
and this defines a bijective mapping
\eq
\label{a17}
\sig :\cU^{(0)}(h)\to\cU^{(0)}(h); U\mapsto \sig(U)\equiv Z
\eqend
with the following properties
\eqa
\label{a18}
\sig(\sig(U))&=& U \quad \forall U\in\cU^{(0)}(h) \nonu
\sig(U)^{-1} &=& \sig(U^{-1}) \quad \forall U\in\cU^{(0)}(h)\nonu
U\in\GB &\Biar& \sig(U)\in \GFn \nonu
U\in \GFn &\Biar& \sig(U)\in\GB
\eqaend
following from \Ref{a15} \cite{R2} (to prove the last two relations,
take the adjoint of \Ref{a15} and use \Ref{a10} and
$T_\pm q_B=q_B T_\pm = \pm T_\pm$).

{\em The theory of boson and fermion BTs can be developed parallely
and on equal footing only if one restricts oneself in the fermion case
to $\GFn$} (see \cite{R1}). This will be done in the following.

{\bf Remark 2.3:} Note that $\GFn$ is not a subgroup of $\GF$.
Fermion BTs with $U\in\GF$ not contained in $\GFn$ play an important
role, e.g.\,  for anomalies and the boson fermion correspondence (see e.g.\
\cite{M} and \cite{CR}).


\sss{(c) Implementers}
The explicit formulas for the implementers $\hGam{U}$, $U\in\GXn$,
$X\in\{B,F\}$, can be found in Ref.\ \cite{R1}. We shall only need
\eq
\label{a19}
\hGam{U}\Om = N(U)\ee{Z(U)_{+-}a^*a^*}\Om
\quad \forall U\in\GXn
\eqend
with $Z(U)\equiv \sig(U)$ \Ref{a16},
\eq
\label{a20}
N(U)=\det{1-\eps_X (Z(U)_{+-})^*Z(U)_{+-}}^{\eps_X/2}
\eqend
a normalization constant ($\det{\cdot}$ is the Fredholm determinant
\cite{S}), where we use the notation
\eq
\label{a20a}
Aa^*a^* = \sum_{n=1}^{\infty} \lam_na^*(f_n)a^*(J g_n)
\eqend
for any operator $A\in\HS$ represented in the standard form \cite{RS1}
\eq
\label{a20b}
A=\sum_{n=1}^{\infty}\lam_nf_n(g_n,\cdot)
\eqend
with $\{f_n\}_{n=1}^{\infty}$ and $\{g_n\}_{n=1}^{\infty}$
orthonormal systems of vectors in $h$ and $\lam_n$ complex numbers.

{\bf Remark 2.4:} Note that the determinant in \Ref{a20} exists if and
only if $Z(U)_{+-}\in\HS$ \cite{S},
and this is the case for all $U\in\GXn$.

{\bf Remark 2.5:} From the estimate \Ref{est} it follows that \Ref{a20}
is well-defined for $A\in\TR$, and the estimate \cite{GL2}
\[
\Norm{Aa^*a^*P_\ell}\leq (\ell+2)\Norm{A}_2\quad\forall\ell\in\N
\]
shows that this definition extends naturally to all $A\in\HS$.

\sss{(c) Current Algebras}
For $A\in\BO$, $\ee{\ii tA}$ is in $\GX$ for all $t\in\R$ if and only
if
\eq
\label{a22}
q_XA^*q_X = A
\eqend
and
\eq
\label{a23}
A_{+-}, A_{-+} \in \HS .
\eqend
We denote a $A\in\BO$ obeying \Ref{a22} as
$X$-self-adjoint\footnote{note that $F$-self-adjoint=self-adjoint},
and as $\gX$ the set of all X-self-adjoint operators obeying
\Ref{a23}. $\gX$ is the Lie algebra of the Lie group $\GX$ (with
$\ii^{-1}\times$commutator as Lie bracket), and it is a Banach algebra
with the norm $\NORM{\cdot}_2$,
\eq
\label{a24}
\NORM{A}_2=\Norm{A_{++}}+\Norm{A_{--}}+\Norm{A_{+-}}_2+\Norm{A_{-+}}_2
\eqend
($\Norm{\cdot}_2$ is the Hilbert-Schmidt norm \cite{RS1}).

For $A=q_XA^*q_X$ in the form \Ref{a20b}, eq.\ \Ref{est}
allows us to define
\eq
\label{a25}
Q(A)\equiv \sum_{n=1}^{\infty}\lam_n\Phi^+(f_n)\Phi(g_n)
\eqend
on $\Df$, and from \Ref{a7a} and \Ref{a7b} we have
\neqa
\ccr{Q(A)}{\Phi^+(f)} &=&\Phi^+(Af)\quad \forall f\in h\\
\ccr{Q(A)}{Q(B)}&=&Q(\ccr{A}{B})\\
Q(A)^*&=&Q(A)
\neqaend
($\ccr{\cdot}{\cdot}$ is the commutator as usual)
for all $A,B\in\HS$.
Thus by defining
\eq
\label{a26}
\dhgam{A}\equiv Q(A)-<\Om,Q(A)\Om> = Q(A)+\eps_X\tra{T_-A}
\eqend
($<\cdot,\cdot>$ is the scalar product in $\cF_X(h)$ and $\tra{\cdot}$
the trace in $h$; the last equality follows from
\Ref{a25}, \Ref{a4}, and \Ref{a2})
we obtain
\eq
\label{a27}
\ccr{\dhgam{A}}{\dhgam{B}}=\dhgam{\ccr{A}{B}}+S(A,B)
\eqend
with $S(A,B)=-\eps_X\tra{T_-\ccr{A}{B}}$ (note that the r.h.s of this is
equal to $Q(\ccr{A}{B}$ for $A,B\in\TR$), or equivalently (by using
the cyclicity of the trace)
\eq
\label{a28}
S(A,B) = -\eps_X\tra{A_{-+}B_{+-}-B_{-+}A_{+-}}
\eqend
which obviously is purely imaginary for all $A,B$ obeying \Ref{a22}.
Moreover,
\eq
\label{a29}
\ccr{\dhgam{A}}{\Phi^+(f)} =\Phi^+(Af)\quad\forall f\in h
\eqend
and $\dhgam{A}$ is essentially self-adjoint implying (we denote the
self-adjoint extension of $\dhgam{A}$ by the same symbol)
\eq
\label{a30}
\ee{\ii t\dhgam{A}}\Phi^+(f)=\Phi^+(\ee{\ii tA}f)
\ee{\ii t\dhgam{A}}\quad\forall t\in\R, f\in h.
\eqend
One can prove the estimate \cite{GL2}
\eq
\label{a31}
\Norm{\dhgam{A}P_\ell}\leq 8\ell\NORM{A}_2 \quad
\forall \ell\in\N
\eqend
showing that the definition of $\dhgam{A}$ naturally extends from
$A=q_XA^*q_X\in\TR$ to $A\in\gX$, and all the relations
\Ref{a27}--\Ref{a31} are valid for all $A,B\in\gX$. The relations
\Ref{a27} provide the (abstract) boson ($X=B$) and fermion ($X=F$)
current algebras, and $S(\cdot,\cdot)$ is the Schwinger term. Due to
the anti-symmetry and the Jacobi identity obeyed by the commutator,
$S(\cdot,\cdot)$ obeys 2-cocycle relations and it is therefor a
(non-trivial) 2-cocycle of the Lie algebra $\gX$, and $\dhgam{\cdot}$
provides a representation of a central extension of $\gX$.

Note that by construction
\eq
\label{phase}
\Produ{\Om}{\dhgam{A}\Om}=0\quad \forall A\in\gX .
\eqend

{\bf Remark 2.6:} The operators $\dhgam{A}$, $A\in\gX$, are unbounded in
general, and (due to \Ref{a31} and $\dhgam{A}P_\ell =
P_{\ell+2}\dhgam{A}P_\ell$ for all $\ell\in\N$)
$\Df$ \Ref{a3} provides a common, dense,
invariant domain of essential self-adjointness for all these
operators \cite{GL2}.

\section{Results}
\label{sec2}
\sss{(a) First Result}
For $U\in\GX$ the defining relation \Ref{a11} determine the
implementer $\hGam{U}$ only up to a phase factor $\in
U(1)\equiv\{\ee{\ii\phi}|0\leq\phi<2\pi\}$, and its unique definition
requires an additional fixing of this phase ambiguity. A convenient
and natural choice for this is \cite{R1}
\eq
\label{a32}
\Produ{\Om}{\hGam{U}\Om} \quad \mbox{real and positiv } \forall
U\in\GXn,
\eqend
and in fact this is the convention used in eqs.\
\Ref{a19} and \Ref{a20}. From the explicit formulas in \cite{R1} one
can easily see that then
\eq
\label{a32a}
\hGam{U}^*=\hGam{U^{-1}}= \hGam{q_X U^* q_X} \quad \forall U\in\GXn .
\eqend

{}From \Ref{a11} it follows that for $U,V\in\GXn$, the unitary operators
$\hGam{U}\hGam{V}$ and $\hGam{UV}$ both implement the same BT
$\alpha_{UV}$, hence they must be equal up to a phase,
\eq
\label{a33}
\hGam{U}\hGam{V}=\chi(U,V)\hGam{UV}
\eqend
with $\chi$ a function $\GX\times\GX\to U(1)$ {\em determined} by
\Ref{a33} and the phase convention used for the implementers.
{}From the assoziativity of the operator
product we conclude that $\chi$ satisfies the relation
\eq
\label{a33a}
\chi(U,V)\chi(UV,W)=\chi(V,W)\chi(U,VW) \quad \forall U,V,W\in\GX,
\eqend
and changing the phase convention for the implementers,
\eq
\label{a34}
\hGam{U}\lrar \beta(U)\hGam{U} \quad \forall U\in\GXn
\eqend
with $\beta:\GXn\mapsto U(1)$ some smooth function, amounts to
changing
\eq
\label{a35}
\chi(U,V)\lrar \chi(U,V) \del \beta(U,V)
\eqend
with
\eq
\label{a36}
\del\beta(U,V)\equiv \f{\beta(U)\beta(V)}{\beta(UV)} \quad \forall
U,V\in\GX
\eqend
satisfying \Ref{a33a} trivially. Eq.\ \Ref{a33a} is a {\em 2-cocycle
relation}, a function $\chi:\GX\times\GX\to U(1)$ satisfying it is a
{\em 2-cocycle}, and a 2-cocycle of the form $\del\beta$ eq.\
\Ref{a36} is a {\em 2-coboundary} of the group $\GX$ \cite{amp2}.

The {\em  first result} of this paper is the explicit formula for the
2-cocycle $\chi$ defined by \Ref{a32} and \Ref{a33}:
\eq
\label{a37}
\chi(U,V)=\left(\f{\det{1+
(V^*_{--})^{-1}(V^*)_{-+}(U^*)_{+-}(U^*_{--})^{-1}}} {\det{1+
(U_{--})^{-1}U_{-+}V_{+-}(V_{--})^{-1}}}
\right)^{\eps_X/2}
\eqend
for all $U,V\in\GXn$.

{\bf Remark 3.1:} Obviously, $V_{+-},U_{-+}\in\HS$ is sufficient
for the existence of the determinants in \Ref{a37} \cite{S}.

{\bf Remark 3.2:} Let $\GXnn$ be the group of all $X$-unitary operators
$U$ on $h$ obeying
\eq
(U -1)\in\TR .
\eqend
Then it is easy to see that for $U,V\in\GXnn\cap\cU^{(0)}(h)$,
\eq
\label{beta}
\chi(U,V) = \del\beta_0(U,V), \quad \beta_0(U)=\left(\f{\det{T_+ +
U_{--}^*}}{\det{T_++U_{--}}} \right)^{\eps_X/2},
\eqend
(cf.\ eq.\ \Ref{cc2}), i.e.\ $\chi$ \Ref{a37} is a trivial 2-cocycle
for for $\GXnn$. However, for $\GXn\ni U\notin \GXnn$, $\beta_0(U)$
\Ref{beta} does not exist in general showing that $\chi$ \Ref{a37} is a
non-trivial 2-cocycle for the group $\GX$.

{\bf Remark 3.3:} Note that in the fermion case $X=F$, eq.\ \Ref{a32} can be
used to determine the phase of the implementers $\hGam{U}$ only for
$U\in\GFn$ (as the l.h.s. of \Ref{a32} is zero
for all other $U\in\GF$ \cite{R1}), and that eq.\ \Ref{a37}
gives the 2-cocycle $\chi$ of the group $\GF$ only {\em locally}, i.e.\ in the
neighborhood $\GFn\subset\GF$ of the identity.

\sss{(b) Second Result}
{}From \Ref{a11} and \Ref{a30} it follows that for $A\in\gX$ and
$t\in\R$, the unitary
operators $\ee{\ii t\dhgam{A}}$ and $\hGam{\ee{\ii tA}}$ both
implement the same BT $\alpha_{\ee{\ii tA}}$, hence they must be equal
up to a phase $\eta(A;t)$,
\eq
\label{a38}
\ee{\ii t\dhgam{A}} = \eta(A;t)\hGam{\ee{\ii tA}}\quad \forall
A\in\gX, t\in\R .
\eqend
The {\em second result} of this paper is the explicit formula for
$\eta(A,t)$ defined by \Ref{a38}, \Ref{a32} and \Ref{a25}, \Ref{a26}:
\eq
\label{a39}
\eta(A;t) = \exp{\left(\ii\int_0^t\dd{r}\phi(\ee{\ii rA},A)\right)} \quad
\forall  A\in\gX
\eqend
for all $t\in\R$ in the boson case $X=B$,  and for all $t\in\R$
such that $\ee{\ii sA}\in\GFn$ for all
$\norm{s}<\norm{t}$ in the fermion case $X=F$, with $\phi$ given by
\eqa
\label{a39a}
\phi(U,A) \equiv
\f{\eps_X}{2}\tra{(U_{--})^{-1}U_{-+}A_{+-} +
(A^*)_{-+}(U^*)_{+-}(U^*_{--})^{-1} } \nonu
\quad \forall U\in\GX , A\in\gX.
\eqaend


\sss{(c) Third Result}
The formula \Ref{a39} can be easily generalized to unitary propagators
$u(t,s)$ generated by a family of $t$-dependent
$X$-self-adjoint operators $A(t)$: Let
\eq
\label{a39b}
A(\cdot):\R\to \gX, t\mapsto A(t)
\eqend
be continuous in the $\NORM{\cdot}_2$-norm. Then
the eq.\
\eqa
\label{a39c}
\f{\partial}{\partial s}u(s,t) = \ii A(s)u(s,t) \nonu
u(t,t) = 1 \quad \forall s,t\in\R
\eqaend
can be solved by $X$-unitary operators $u(s,t)$ obeying
\eq
\label{a41}
u(r,s)u(s,t) = u(r,t) \quad \forall r,s,t \in\R
\eqend
(see e.g.\ \cite{RS2}, Section X.12).
As a {\em third result} of this paper we show that
\eq
\label{a42}
u(s,t)\in\GX \quad \forall s,t\in\R,
\eqend
and that the second quantized version of \Ref{a39c}
\eqa
\label{a43}
\f{\partial}{\partial s}U(s,t) = \ii \dhgam{A(s)}U(s,t) \nonu
U(t,t) = 1 \quad \forall s,t\in\R
\eqaend
can be solved by unitary operators $U(s,t)$ on $\cF_X(h)$;
moreover
\eq
\label{a45}
U(s,t) =
\eta(A(\cdot),s,t)\hGam{u(s,t)} \quad
\forall (s,t)\in I^2
\eqend
with
\eq
\label{a45a}
\eta(A(\cdot),s,t) = \exp{\left( \ii\int_t^s\dd{r}\phi(u(s,r),A(r))\right)}
\eqend
and $\phi$ eq.\ \Ref{a39a}; the domain of validity for this is $I^2=\R^2$
in the boson case $X=B$, and $I^2$ the set of all $(s,t)\in\R^2$ such that
$u(s,t)\in\GXn$ in the fermion case $X=F$.

{\bf Remark 3.4:} Though the second result is a special case of the third
one, we prefer to state and prove it independently: the non-trivial
part of the later is the existence of the unitary propagator $U(s,t)$
(which is trivial for $t$-intependent generators $A$), whereas
the proof of eqs.\  \Ref{a45}--\Ref{a45a} can be given by a straightforward
extension of the one for \Ref{a38}--\Ref{a39}.

\section{Proofs}
\label{sec3}
\sss{(a) Proof of the First Result}
We first proof the following Lemma:
\subsubsection*{Lemma:} Let $A_{+-},B_{+-}\in\HS$. Then
\eq
\label{bb1}
\Produ{\ee{A_{+-}a^*a^*}\Om}{\ee{B_{+-}a^*a^*}\Om}=
\det{1-\eps_X(A_{+-})^*B_{+-}}^{-\eps_X}.
\eqend

{\em Proof of the Lemma:} By assumption,
\eq
\label{b3}
T\equiv (A_{+-})^*B_{+-}\in\TR ,
\eqend
and we assume at first that $\Norm{A_{+-}},\Norm{B_{+-}}<1$.  Similarly
as in \cite{R1} one can easily show then that
\eq
\label{bb2a}
\mbox{l.h.s. of \Ref{bb1}}=\sum_{n=0}^{\infty} a_n^{(X)}
\eqend
with
\eq
\label{bb2b}
a_n^{(X)}= \f{1}{n!}\sum_{m_1,m_2,\cdots m_n =1}^{\infty}
\sum_{\pi\in\cP_n}\sign(\pi)^{\dif{X}}
\prod_{i=1}^n \produ{e_{m_i}}{Te_{m_{\pi(i)}}}
\eqend
$\{e_{m}\}_{m=0}^\infty$ a complete, orthonormal basis in $T_-h$,
$\cP_n$ the set of all permutation of $\{1,2,\cdots n\}$, and
$\sign(\pi)=1(-1)$ for even (odd) permutations $\pi\in\cP_n$.

Now obviously
\eqa
\label{yyy}
\prod_{i=1}^n \produ{e_{m_i}}{Te_{m_{\pi(i)}}} =
\produ{e_{\tilde{m}_1}}{Te_{\tilde{m}_1}}
\produ{e_{\tilde{m}_2}}{Te_{\tilde{m}_2}}
\nonu \times \cdots
\produ{e_{\tilde{m}_{N_1}}}{Te_{\tilde{m}_{N_1}}}
\produ{e_{\tilde{m}_{N_1+1}}}{Te_{\tilde{m}_{N_1+2}}}
\nonu \times
\produ{e_{\tilde{m}_{N_1+2}}}{Te_{\tilde{m}_{N_1+1}}}
\produ{e_{\tilde{m}_{N_1+3}}}{Te_{\tilde{m}_{N_1+4}}}
\produ{e_{\tilde{m}_{N_1+4}}}{Te_{\tilde{m}_{N_1+3}}}
\cdots \nonu \times
\produ{e_{\tilde{m}_{N_1+2N_2-1}}}{Te_{\tilde{m}_{N_1+2N_2}}}
\produ{e_{\tilde{m}_{N_1+2N_2}}}{Te_{\tilde{m}_{N_1+2N_2-1}}}
\produ{e_{\tilde{m}_{N_1+2N_2+1}}}{Te_{\tilde{m}_{N_1+2N_2+2}}}
\nonu \times
\produ{e_{\tilde{m}_{N_1+2N_2+2}}}{Te_{\tilde{m}_{N_1+2N_2+3}}}
\produ{e_{\tilde{m}_{N_1+2N_2+3}}}{Te_{\tilde{m}_{N_1+2N_2+1}}}
\cdots , \qquad
\eqaend
with
$(\tilde{m}_1,\tilde{m_2},\cdots,\tilde{m}_n) =
(m_{\sig(1)},m_{\sig(2)},\cdots,m_{\sig(n)})$ for some $\sig\in\cP_n$
and non-negativ integers $N_1,N_2,\cdots N_n$ obeying
\eq
\label{sum}
\sum_{\alpha=1}^n \alpha N_\alpha = n
\eqend
and determined by $\pi\in\cP_n$. Hence we have
\neqa
\sum_{m_1,m_2,\cdots m_n =1}^{\infty}
\prod_{i=1}^n \produ{e_{m_i}}{Te_{m_{\pi(i)}}} \\
= \tra{T}^{N_1}\tra{T^2}^{N_2}\cdots \tra{T^n}^{N_n} =
\prod_{\alpha=1}^n \tra{T^\alpha}^{N_{\alpha}} .
\neqaend
{}From \Ref{yyy} we can deduce that
\[
\sign(\pi) = \prod_{\alpha=1}^n (-)^{(\alpha -1)N_\alpha} ,
\]
and we can write
\eq
\label{bb4}
a_n^{(X)} = \f{1}{n!}\mbox{$\sum_{N_1,N_2,\cdots}^{(n)}$}
\, K_n(N_1,N_2,\cdots N_n) \prod_{\alpha
=1}^n (-)^{\dif{X}(\alpha -1)N_{\alpha}}\tra{T^\alpha}^{N_{\alpha}}
\eqend
with $\sum_{N_1,N_2,\cdots}^{(n)}$ the sum over
all non-negative integers $N_1,N_2,\cdots$ obeying \Ref{sum}, and
$K_n(N_1,N_2,\cdots N_n)$ denoting the number of permutations
$\pi\in\cP_n$ leading to a term containing
$\prod_{\alpha=1}^n\tra{T^\alpha}^{N_\alpha}$.
In the following we determine these numbers by simple combinatorics.

There are $n$ numbers $(\tilde{m}_1,\tilde{m_2},\cdots,\tilde{m}_n)$,
hence we have $n$ possibilities to choose $\tilde{m}_1$,
$(n-1)$ possibilities for $\tilde{m}_2$, $\cdots$, $(n-N_1+1)$
possibilities to choose $\tilde{m}_{N_1}$; however
$N_1!=N_1(N_1-1)\cdots 1$ of these $n(n-1)\cdots (n-N_1+1)$ choices
are equal as permutations of
$(\tilde{m}_1,\tilde{m_2},\cdots,\tilde{m}_{N_1})$ have to be
identified. Hence we have
\[
\f{n!}{(n-N_1)!N_1!}
\]
different choices for $(\tilde{m}_1,\tilde{m_2},\cdots,\tilde{m}_{N_1})$
for producing a term containing $\tra{T}^{N_1}$. In order to
produce a term containing in addition
$\tra{T^2}^{N_2}$, we have $(n-N_1)$ possibilities left to choose
$\tilde{m}_{N_1+1}$, $\cdots$, $(n-N_1-2N_2+1)$ possibilities to
choose $\tilde{m}_{N_1+2N_2}$. Again we have to identify the choices
obtained by permuting
$(\tilde{m}_{N_1+1},\cdots,\tilde{m}_{N_1+2N_2})$; moreover (due to
the cyclicity of the trace), we must
identify the choices which differ only by changing
$\tilde{m}_{N_1+1}$ and $\tilde{m}_{N_1+2}$, $\cdots$, or
$\tilde{m}_{N_1+2N_2-1}$ and
$\tilde{m}_{N_1+2N_2}$; all together we have
\[
\f{n!}{(n-N_1)!N_1!}\f{(n-N_1)!}{(n-N_1-2N_2)!N_2!2^{N_2}}
\]
different
choices for $(\tilde{m}_1,\tilde{m_2},\cdots,\tilde{m}_{N_1+2N_2})$
leading to a term containing
$\tra{T}^{N_1}\tra{T^2}^{N_2}$.
Continuing these considerations, we arrive at the conclusion that there
are
\eqa
\label{bb5}
\f{n!}{(n-N_1)!N_1!}\f{(n-N_1)!}{(n-N_1-2N_2)!N_2!2^{N_2}}
\f{(n-N_1-2N_2)!}{(n-N_1-2N_2-3N_3)!N_3!3^{N_3}}\cdots
\nonu = n!\prod_{\alpha=1}^n\f{1}{N_\alpha
!}\left(\f{1}{\alpha}\right)^{N_\alpha} = K_n(N_1,N_2,\cdots N_n)
\qquad
\eqaend
different permutations $\pi\in\cP_n$ leading to a $\prod_{\alpha
=1}^n\tra{T^\alpha}^{N_{\alpha}}$-term.
(As a simple check, one can easily convince oneselves that
\[
\mbox{$\sum_{N_1,N_2,\cdots}^{(n)}$}
\, K_n(N_1,N_2,\cdots N_n)=n!.\mbox{)}
\]

With that we obtain
\[
a_n^{(X)}=\mbox{$\sum_{N_1,N_2,\cdots}^{(n)}$}\prod_{\alpha=1}^n
\eps_X^{(\alpha-1)N_\alpha}
\f{\tra{T^\alpha}^{N_\alpha}}{N_\alpha !\alpha^{N_\alpha}}
\]
leading to
\neqa
\sum_{n=0}^\infty a_n^{(X)} = \sum_{N_1,N_2,\cdots=0}^\infty
\prod_{\alpha=1}^\infty \f{1}{N_\alpha !}\left(\eps_X^{\alpha -1}
\f{\tra{T^\alpha}}{\alpha} \right)^{N_\alpha}
= \prod_{\alpha=1}^\infty \exp{\left(\eps_X^{\alpha -1}
\f{\tra{T^\alpha}}{\alpha} \right)} \\
= \exp{\left(-\eps_X\tra{\log{(1-\eps_X T)}}\right)} =
\det{1-\eps_X T)}^{-\eps_X}
\neqaend
where we freely interchanged infinite sums and products (which is
allowed as everything converges absolutely for all $T\in\TR$,
$\Norm{T}<1$).

The validity of \Ref{bb1} for general $T$ \Ref{b3} follows from the
observation that due to the considerations above,
\[
\Produ{\ee{\bar{z}_1A_{+-}a^*a^*}\Om}{\ee{z_2B_{+-}a^*a^*}\Om}=
\det{1-\eps_X(\bar{z}_1A_{+-})^*z_2B_{+-}}^{-\eps_X}, \quad z\in\C
\]
is valid for $\norm{z_1}<1/\Norm{A_{+-}}$,
$\norm{z_2}<1/\Norm{B_{+-}}$, and both sides of this eq.\ are analytic
have a unique analytic extension to $z_1=z_2=1$.

{\bf Remark 4.1:} In the fermion case $X=F$, eq.\ \Ref{bb1}
was given by Ruijsenaars \cite{R3}.

{\em Proof of \Ref{a37}:} With \Ref{a33} we have (cf.\ \Ref{a32a})
\[
\Produ{\hGam{U^{-1}}\Om}{\hGam{V}\Om} =
\chi(U,V)\Produ{\Om}{\hGam{UV}\Om},
\]
and with \Ref{a19}
\[
\chi(U,V)= \f{N(U^{-1})N(V)}{N(UV)}\cE(U,V)
\]
where
\neqa
\cE(U,V) =
\Produ{\ee{Z(U^{-1})_{+-}a^*a^*}\Om}{\ee{Z(V)_{+-}a^*a^*}\Om} = \\
\det{1-\eps_X(Z(U^{-1})_{+-})^* Z(V)_{+-}}^{-\eps_X}
\neqaend
(we used the Lemma above).
Now obviously for all $U\in\GXn$ (cf.\ \Ref{a16})
\[
(Z(U^{-1})_{+-})^* = \eps_X Z(U)_{-+}
\]
hence with \Ref{a16}
\neqa
1-\eps_X(Z(U^{-1})_{+-})^* Z(V)_{+-} =
1-(U_{--})^{-1}U_{-+}V_{+-}(V_{--})^{-1} \\ =
T_+ + (U_{--})^{-1}(UV)_{--}(V_{--})^{-1} ,
\neqaend
and we can write
\[
\cE(U,V)= \detp{(U_{--})^{-1}(UV)_{--}(V_{--})^{-1}}^{-\eps_X}
\]
with
\[
\detp{\cdots} \equiv \det{T_+ + \cdots}
\]
the determinant on $T_-h$. Similarly one obtains from \Ref{a20}
\[
N(U)=\detp{(U_{--}^*)^{-1}(U_{--})^{-1}}^{\eps_X/2}
\]
(which can also be deduced from $N(U)^{-2}=\cE(U^{-1},U)$ and
$(U^{-1})_{--} = U^*_{--}$),
and by simple properties of determinants this results in
\eq
\label{cc2}
\chi(U,V)= \left(\f{\detp{(V^*_{--})^{-1}(V^*U^*)_{--}
(U^*_{--})^{-1}}}{\detp{(U_{--})^{-1}(UV)_{--}
(V_{--})^{-1}}}\right)^{\eps_X/2} .
\eqend
Using
\[
(UV)_{--}= U_{--}V_{--}+U_{-+}V_{+-}
\]
and its adjoint we obtain \Ref{a37}.

{\bf Remark 4.2:} Note that we can write \Ref{cc2} also as
\eq
\label{cc2mod}
\chi(U,V)=\left(\f{\detp{V_{--}(V^*U^*)_{--}U_{--}}}{
\detp{U_{--}^*(UV)_{--}V_{--}^*}}\right)^{\eps_X/2}.
\eqend

\sss{(b) Proof of the Second Result}
Let $A\in\gX$ and $s,t,\epsilon\in\R$. We write for simplicity
\eq
\chi(\ee{\ii sA},\ee{\ii tA}) \equiv \chi'(s,t) .
\eqend
Due to the cocycle relation \Ref{a33} we have
\[
\chi'(s,t)=\f{\chi'(t,\epsilon)}{\chi'(s+t,\epsilon)}
\chi'(s,t+\epsilon),
\]
and by iteration
\[
\chi'(s,t) = \left(\prod_{\nu=0}^{N-1}\f{\chi'(t+\nu
\epsilon,\epsilon)}{\chi'(s+t+\nu
\epsilon,\epsilon)}\right)\chi'(s,t+N\epsilon)
\]
valid for arbitrary $N\in\N$. Choosing $t+N\epsilon=0$ and noting that
$\chi'(s,0)=1$, we obtain the formula
\eq
\chi'(s,t) = \left(\prod_{\nu=0}^{N-1}\f{\chi'(t-\nu t/N ,-t/N)}
{\chi'(s+t-\nu t/N,-t/N)}\right) \quad \forall N\in\N .
\eqend
For $s,t$ such that $\ee{\ii sA},\ee{\ii tA}, \ee{\ii(s+t)A} \in\GXn$
we have
\eq
\log{(\chi'(s,-t/N))} = -c(s)t/N + \OO((t/N)^2), \quad
c(s)=\f{\dd}{\dd t}\log{(\chi'(s,t))}\restr{t=0}
\eqend
(note that $\chi'(s,t)$ is continuously differentiable),
hence
\neqa
\log{(\chi'(s,t))} = \sum_{\nu=0}^{N-1}\left(c(t+s-\nu t/N)
- c(t-\nu t/N) \right)t/N + \OO(t/N) \\
= \int_{0}^t \dd{r}\left(c(s+r)-c(r) \right) + \OO(t/N)
\\  =
\log{(\eta'(t+s))} - \log{(\eta'(s))} - \log{(\eta'(t))} + \OO(t/N)
\neqaend
with
\eq
\log{(\eta'(s))} = \int_0^s\dd{r}c(r).
\eqend
In the limit $N\to\infty$ we obtain
\eq
\chi'(s,t) = \left(\f{\eta'(s)\eta'(t)}{\eta'(s+t)}\right)^{-1},
\eqend
equivalent to
\[
\hat{\Gamma}'(t)\equiv \eta'(t)\hGam{\ee{\ii tA}}
\]
obeying
\[
\hat{\Gamma}'(s)\hat{\Gamma}'(t)=\hat{\Gamma}'(s+t).
\]
Moreover, a simple calculation yields
\[
\f{\dd}{\dd\ii t}\Produ{\Om}{\hat{\Gamma}'(t)\Om}\restr{t=0}=0
\]
(cf. \Ref{phase}).
{}From the uniqueness of an implementer of a Bogoliubov transformation
with a fixed phase we can conclude thus that
\[
\hat{\Gamma}'(t) = \ee{\ii t\dhgam{A}}
\]
yielding \Ref{a38} with
\eq
\eta(A;t)= \eta'(t).
\eqend

Now
\eq
c(s) = \ii \phi(\ee{\ii sA},A)
\eqend
with
\eq
\phi(U,A)\equiv\f{\dd}{\dd \ii t}\log{( \chi(U,\ee{\ii tA}))}\restr{t=0} ,
\eqend
and with \Ref{cc2mod} we obtain \Ref{a39a}.

\sss{(c) Proof of the Third  Result}
Let $s,t\in\R$.
As $A(\cdot)$ \Ref{a39b} is continuous in the $\NORM{\cdot}_2$-norm,
we have from the principle of uniform boundedness \cite{RS2} that
\eq
\label{x1}
\NORM{A(r)}_2 < \alpha(s,t) \quad \forall r\in[s,t]
\eqend
($[s,t]\subset\R$ denotes the closed interval inbetween $s$ and $t$)
for some finite $\alpha(s,t)$, and obviously $u(r,t)$ \Ref{a39c} is
uniformly bounded in the operator norm for all $r\in [s,t]$ as well.
{}From \Ref{a39c} we obtain
\[
u(s,t)=1+\int_t^s\dd{r}\ii A(r) u(r,t),
\]
hence
\[
\NORM{u(s,t)}_2 \leq 1 + |s-t|\alpha(s,t)
\sup_{r\in[s,t]}\Norm{u(r,t)}
\]
proving \Ref{a42}.

{}From \Ref{a31} and \Ref{a25}, \Ref{a26} we have
\eqa
\label{x2}
\Norm{\dhgam{A(r)}P_\ell} &\leq & 8\ell \alpha(s,t) \nonu
\dhgam{A(r)}P_\ell &=& P_{\ell +2}\dhgam{A(r)}P_\ell \quad \forall \ell
\in\N , r\in[s,t] .
\eqaend
Eq.\ \Ref{a43} is formally equivalent to
\[
U(s,t)=1+\int_t^s\dd{r}\, \ii \dhgam{A(r)} U(r,t),
\]
suggesting to define $U(s,t)$ as power series on $\cD^f(h)$
\eq
\label{x3}
U(s,t)=1+\sum_{n=1}^\infty R_n(s,t)
\eqend
with
\eq
\label{x4}
R_n(s,t) =\int_t^s\dd{r_1}\ii\dhgam{A(r_1)}
\int_t^{r_1}\dd{r_2}\ii\dhgam{A(r_2)}\cdots
\int_t^{r_n}\dd{r_{n-1}}\ii\dhgam{A(r_{n-1})}.
\eqend
Indeed, with \Ref{x2} we can estimate
\neqa
\Norm{R_n(s,t)P_\ell} \leq |\int_t^{s}\dd{r_1}\int_t^{r_1}\dd{r_2}
\cdots\int_t^{r_n}\dd{r_{n-1}} \\ \times
\Norm{\dhgam{A(r_1)}P_{\ell +2n-2}}
\Norm{\dhgam{A(r_2)}P_{\ell +2n-4}}\Norm{\dhgam{A(r_n)}P_{\ell}}|
\\ \leq (8\alpha(s,t))^n\ell (\ell +2) \cdots (\ell +2n -2)\f{|s-t|^n}{n!},
\neqaend
and with the ratio test for power series we conclude that \Ref{x3},
\Ref{x4} is well-defined on $\cD^f(h)$ for $16\alpha(s,t)|s-t|<1$.
Then for sufficiently small $|s-t|$, $|r-t|$, $|r-s|$,
\Ref{x3} and \Ref{x4} imply that
\eq
\label{x5}
U(r,s)U(s,t)=U(r,t)
\eqend
and
\eq
U(s,t)^*=U(t,s)
\eqend
(we recall that all $\dhgam{A(r)}$, $r\in\R$, are essentially
self-adjoint on $\cD^f(h)$)
showing that $U(s,t)$ can be uniquely extended to a unitary operator
on $\cF_X(h)$ such that \Ref{x5} remains true. Then we can use
\Ref{x5} to extend the definition of $U(s,t)$ to all $s,t\in\R$, and
it is easy to see that these satisfy \Ref{a43}.

Having established the existence of $U(s,t)$ \Ref{a43}, the validity
of \Ref{a45} and \Ref{a45a} can be proved similarly as
\Ref{a38}--\Ref{a39a}: Defining
\eq
\chi'_s(r,t)\equiv \chi(u(r,s),u(s,t))
\eqend
we deduce from the cocycle relation \Ref{a37} that
\neqa
\chi'_s(r,t) =
\f{\chi'_t(s,t+\epsilon)}{\chi'_t(r,t+\epsilon)} \chi'_s(r,t+\epsilon)
\\ = \left(\prod_{\nu=1}^{N}
\f{\chi'_{t+(\nu -1)\epsilon} (s,t+\nu \epsilon)}
{\chi'_{t+(\nu -1)\epsilon}(r,t+\nu\epsilon)}\right)
\chi'_s(r,t+N\epsilon) ,
\neqaend
and with
\[
t+N\epsilon =s, \quad \chi'_s(r,s)=1
\]
we obtain similarly as above
\eq
\chi'_s(r,t) = \left(\f{\eta'_s(t)\eta'_r(s)}{\eta'_r(t)}\right)^{-1}
\eqend
with
\eq
\log{(\eta'_s(t))}= -\int_t^s\dd{q}c(s,q),
\eqend
\eqa
c(s,q)=\f{\partial}{\partial r} \log{(\chi(u(s,q),u(q,r))}\restr{r=q}
= -\ii\phi(u(s,q),A(q))
\eqaend
and $\phi$ \Ref{a39a}, proving \Ref{a45}, \Ref{a45a} by the same
argument as above (note that
\[
\f{\dd}{\dd \ii
s}\Produ{\Om}{\eta'_s(t)\hat{\Gamma}(u(s,t))\Om}\restr{s=t} = 0
\]
(cf. \Ref{phase})).

\section{Final Remarks}
\label{sec4}

\sss{Remark 5.1} It is easy to see that the current algebra \Ref{a27} is in
fact the Lie algebra version of \Ref{a33}. Indeed, it follows from \Ref{a38}
and $\dd{\eta(A,t)}/\dd{t}\restr{t=0}=0$ that
\eq
\dhgam{A}=\f{\dd}{\dd \ii t} \hGam{\ee{\ii tA}}\restr{t=0} \quad
\forall A\in\gX ,
\eqend
hence
\eqa
\ccr{\dhgam{A}}{\dhgam{B}} = \f{\dd}{\dd \ii s}\f{\dd}{\dd \ii t}
\hGam{\ee{\ii sA}}\hGam{\ee{\ii tB}}\hGam{\ee{-\ii sA}}\hGam{\ee{-\ii tB}}
\restr{s=t=0} \nonu = \dhgam{\ccr{A}{B}}+S(A,B) \quad \forall
A,B\in\gX
\eqaend
with
\eq
S(A,B) = \f{\dd}{\dd \ii s}\f{\dd}{\dd \ii t} \chi(\ee{\ii sA},\ee{\ii
tB}) \chi(\ee{-\ii sA},\ee{-\ii tB}) \chi(\ee{\ii sA}\ee{\ii tB},
\ee{-\ii sA}\ee{-\ii tB})\restr{s=t=0} ;
\eqend
with $\chi$ \Ref{a37} this can be shown to coincide with \Ref{a28}.

\sss{Remark 5.2} For simplicity, we restricted our discussion in this
paper to currents $\dhgam{A}$ with {\em bounded} operators $A$ on
$h$. However, it is straightforward to extend all our results from the
Lie algebra $\gX$ to certain Lie algebras $\gXu$ of {\em unbounded}
operators $A$ on $h$ with a common, dense, invariant domain of
definition which are naturally associated with some given self-adjoint
operator $H$ on $h$ \cite{GL2}, and with the results of \cite{GL2} it
is straightforward to show that \Ref{a38}, \Ref{a39} and \Ref{a45},
\Ref{a45a}  hold for all $A\in\gXu$ and mappings
\[
A(\cdot): \R\to \gXu, t\mapsto A(t)
\]
continuous in the natural topology of $\gXu$ \cite{GL2}, respectively.
Hence in general one can use these formulas for unbounded operators as
well.

\sss{Remark 5.3} As an application of \Ref{a45}, we mention
that it can be used to construct the time evolution $U(s,t)$ on the
second quantized level if the time evolution $u(s,t)$ on the
1-particle level is given and $u(s,t)\in\GXn$.  For example, the later
condition is fulfilled for bosons or fermions in external Yang-Mills
fields in $(1+1)$- (but not in higher) dimensions (in the fermion
case, \Ref{a45} is valid only as long as $u(s,t)$ does not produce
level crossing).


\newpage
\bc
\section*{Acknowledgement}
\ec
I would like to thank S. N. M. Ruijsenaars for the hospitality at the
CWI in Amsterdam where part of this work was done, and also for
helpful and interesting discussion.
Financial support of the ``Bundeswirtschaftkammer''
of Austria is appreciated.
\vspace*{1cm}


\end{document}